\documentclass[namedreferences]{solarphysics}
\usepackage[hyperref,optionalrh,solaromanenum]{spr-sola-addons} 
\usepackage{graphicx}                    
\usepackage{color}                       
\usepackage{breakurl}                         

\ifx\urlurl    \undefined
\def\urlurl#1{\href{http://#1}{\textsf{#1}}}\fi

\newcommand{\ion}[2]{{#1} {\sc #2}}
\newcommand{\lam}{$\lambda$}
\newcommand{\arcsec}{$^{\prime\prime}$}
\newcommand{\as}{$^{\prime\prime}$}
\newcommand{\ecss}{erg\,cm$^{-2}$\,s$^{-1}$\,sr$^{-1}$}
\newcommand{\hinode}{\textit{Hinode}}

\chardef\us=`\_

\begin{document}

\begin{article}

\begin{opening}

\title{Properties of \emph{EUV Imaging Spectrometer} (EIS) Slot Observations\footnote{Software and data files used for creating figures for this article are available at \urlurl{github.com/pryoung/papers/tree/main/2022\_eis\_slot.}}}

%
\author[addressref={aff1,aff2},corref,email={peter.r.young@nasa.gov}]{\inits{P.R.}\fnm{Peter R.}~\lnm{Young}\orcid{0000-0001-9034-2925}}
\author[addressref={aff3}]{\inits{I.}\fnm{Ignacio}~\lnm{Ugarte-Urra}\orcid{0000-0001-5503-0491}}

\address[id=aff1]{NASA Goddard Space Flight Center, Code 671, Heliophysics Science Division, Greenbelt, MD 20771, USA}
\address[id=aff2]{Department of Mathematics, Physics and Electrical Engineering, Northumbria University, Newcastle upon Tyne, UK}
\address[id=aff3]{Space Science Division, Naval Research Laboratory, Washington, DC 20375, USA}

\runningauthor{P.R.~Young \& I.~Ugarte-Urra}
\runningtitle{EIS Slot Observations}

%
\runningauthor{}
\runningtitle{}


\begin{abstract}
The \textit{Extreme ultraviolet Imaging Spectrometer} (EIS) on board the \hinode\ spacecraft has been operating since 2006, returning high resolution data in the 170--212 and 246--292\,\AA\ wavelength regions. EIS has four slit options, with the narrow 1\as\ and 2\as\ slits used for spectroscopy and the wide 40\as\ and 266\as\ slits used for monochromatic imaging. In this article several properties of the 40\as\ slit (or slot) are measured using the \ion{Fe}{xii} 195.12~\AA\ line, which is formed at 1.5~MK. The projected width of the slot on the detector shows a small variation along the slit with an average value of 40.949\as. The slot image is tilted on the detector and a quadratic formula is provided to describe the tilt. The tilt corresponds to four pixels on the detector and the slot centroid is offset mostly to the right (longer wavelengths) of the 1\as\ slit by up to four pixels. Measurement of the intensity decrease at the edge of the slot leads to an estimate of the spatial resolution of the images in the x-direction. The resolution varies quadratically along the slot, with a minimum value of 2.9\as\ close to the detector center. Intensities measured from the slot images are found to be on average 14\%\ higher than those measured from the 1\as\ slit at the same spatial location. Background subtraction is necessary to derive accurate intensities in quiet Sun and coronal hole regions.  Prescriptions for deriving accurate slot intensities for different types of slot datasets are presented.
\end{abstract}

%

\end{opening}

\section{Introduction}

The imaging slit spectrometer has been an important tool for studying the solar atmosphere at ultraviolet wavelengths, with routine observations beginning with the \emph{Coronal Diagnostic Spectrometer} \citep{1995SoPh..162..233H}, the \emph{Solar Ultraviolet Measurements of Emitted Radiation} \citep{1995SoPh..162..189W} and the \emph{Ultraviolet Coronagraph Spectrometer} \citep{1995SoPh..162..313K}, all launched on the \emph{Solar and Heliospheric Observatory} in 1995. The standard mode of operation of these instruments is to image the Sun through a narrow slit with a width of size close to the spectral resolving element of the instrument. The resulting image on the detector shows a set of parallel lines whose positions are the wavelengths of the atomic transitions in the spectrum. The variation of intensity along a line is due to the varying emission of solar features along the slit direction.
Images of the Sun can be obtained by scanning the solar image across the slit, which builds up a two-dimensional image exposure-by-exposure in a procedure called ``rastering". This can be a time-consuming process and a scan of an active region may take tens of minutes.

High cadence extreme ultraviolet (EUV) imaging from first the \textit{EUV Imaging Telescope} \citep{1995SoPh..162..291D} and more recently the \textit{Atmospheric Imaging Assembly} \citep[AIA:][]{2012SoPh..275...17L} have shown that the EUV corona can be highly dynamic. For example, \citet{2021A&A...647A.159C} studied EUV bursts in quiet Sun and a coronal hole and found lifetimes down to the resolution of AIA (24~s). It is thus clear that a single raster image from a spectrometer cannot capture such evolution, yet the plasma diagnostics obtained from a spectrometer are still desirable for understanding the plasma properties.


A compromise solution comes from taking spectroscopic measurements with a wide slit, often referred to as a slot. The detector image will then, in place of the spectral lines, display two-dimensional images at the locations of each of the atomic transitions. By repeated exposures at a spatial location, high-cadence image data can be obtained with the spectrometer. Alternatively, rasters can be performed for larger spatial regions and an image of an entire active region can be obtained in minutes. The disadvantage of this mode is that the density of atomic transitions in the ultraviolet spectrum is high and so images in nearby lines will overlap, complicating the analysis. Strong, relatively isolated lines can be used with confidence, however. Image ratios of density or temperature-sensitive diagnostic atomic transitions can potentially be used to yield high-cadence plasma parameters not possible with EUV imagers.

The present article discusses properties of the 40\arcsec\ slot that is a component  of the \emph{EUV Imaging Spectrometer} \citep[EIS:][]{2007SoPh..243...19C} on the \emph{Hinode} spacecraft. It complements the 1\arcsec\ and 2\arcsec\ slits that are used for spectroscopy. \emph{Hinode} was launched in 2006 and the first use of slot data for science analysis was by \citet{2007PASJ...59S.751C}, who analyzed three-step rasters with a cadence of 3~min to measure the properties of a coronal hole jet in a wide range of emission lines. 

A novel application of slot data was performed by \citet{2008SoPh..252..283I}, who searched for emission that extended beyond the left and right edges of the slot. Such emission is caused by large Doppler shifts of features at the edges of the slot. Velocities of more than 100\,km\,s$^{-1}$ were found in lines such as \ion{He}{ii} \lam256.32 and \ion{Fe}{xv} \lam284.16.

\citet{2009ApJ...695..642U} used slot rasters to study the heating and cooling of active region loops over the temperature range 0.4 to 2.5~MK at a cadence of 70~s. Further loop studies were performed by 
\citet{2011ApJ...730...37U} and \citet{2011ApJ...727...58W} who used slot rasters to study the dynamics of active region outflow regions and fan loops, particularly focusing on \ion{Si}{vii} \lam275.36 and \ion{Fe}{xii} \lam195.12 formed at 0.6~MK and 1.5~MK, respectively. \ion{Si}{vii} showed apparent inflows in the fan loops in the slot movies while \ion{Fe}{xii} showed apparent outflows; both types of motion were confirmed through narrow slit Doppler maps from the same regions.

Time series observations with the slot have been used to study waves and oscillations in various types of solar structure. \citet{2009A&A...499L..29B} identified slow magnetosonic waves above the limb in coronal holes using \ion{Fe}{xii} \lam195.12, with periods in the range 10 to 30~min and propagation speeds around 125\,km\,s$^{-1}$. Oscillations in \lam195.12 were also reported by \citet{2009A&A...494..355O} from an active region slot dataset with a high 3.1~s cadence. Another \lam195.12 active region loop dataset, with 20~s cadence, yielded evidence for standing slow acoustic oscillations at the looptop and one footpoint \citep{2010NewA...15....8S}. \ion{He}{ii} \lam256.32, \ion{Fe}{xii} \lam195.12 and \ion{Fe}{xv} \lam284.16 slot images from a quiet Sun bright point were analyzed by \citet{2010MNRAS.405.2317S} to identify 5~min oscillations in \ion{He}{ii} and \ion{Fe}{xii} lines, but not \ion{Fe}{xv}. The cadence was 32~s.

\citet{2018SoPh..293...56K} used a sequence of slot images at a fixed location in the quiet Sun to study morphology at different temperatures in relation to the magnetic field, and they identified two jets in the \ion{He}{ii} \lam256.3 line.

\citet{2017ApJ...842...58H,2020SoPh..295...34H} have demonstrated that the  266\arcsec\ slot data can be used to measure Doppler shifts and identify the hot \ion{Fe}{xxiv} \lam255.1 emission (formed at 18~MK) line in flares.

Usage of the EIS slot reduced following the launch of AIA on the \emph{Solar Dynamics Observatory} in 2010 as this instrument continuously images the entire solar disk in seven different EUV wavelengths at 12~s cadence and with a higher spatial resolution than EIS. The AIA data can, however, be compromised by uncertainty over the species that are contributing to the EUV channels' bandpasses \citep[e.g.,][]{2012A&A...540A..24B}. Hence, there is still value in high cadence slot imaging with EIS where, at least for strong, unblended lines, one can expect monochromatic images. In addition, there has been renewed interest in slot or slitless spectrometers \citep{2021FrASS...8...50Y} due to the potential of applying deconvolution codes to resolve the spectral and spatial dimensions and thus recover emission line profiles \citep{2019ApJ...882...12W}.

As of October 2021, slot observations comprise  9.2\,\%\ of all EIS observations in terms of observation duration, with  205~days of accumulated data. Given the relatively low number of papers that use slot data, as described above, this is  a large dataset that mostly remains unexplored.
The present article determines technical parameters for the EIS slot that will be valuable for scientists who wish to exploit the data. The particular questions addressed are:
\begin{enumerate}
\item What is the measured width of the slot?
\item What is the tilt of the slot on the detector?
\item What is the spatial resolution of the slot?
\item Are the intensities measured with the slot compatible with the narrow slits?
\end{enumerate}

Section~\ref{sect.overview} provides an overview of EIS and the slot data. Section~\ref{sect.twsr} addresses the first three questions listed above, and Section~\ref{sect.int} the fourth question. In each case we focus on the strong \ion{Fe}{xii} 195.12~\AA\ emission line. Section~\ref{sect.pres} provides a prescription for deriving intensities from slot data, and Section~\ref{sect.summary} summarizes the results of the article.

\section{Overview of EIS and the Slot Data}\label{sect.overview}

EIS is described in \citet{2007SoPh..243...19C} and it features a short wavelength (SW) channel from 170 to 212\,\AA\ and a long wavelength (LW) channel from 246 to 292\,\AA. Spectral resolution is  3000\,--\,4000 and the spatial resolution is 3--4\arcsec. Pixels on the detector correspond to 1\arcsec\ $\times$ 1\arcsec. A slit/slot mechanism is placed between the primary mirror and the grating, limiting the image field-of-view that reaches the grating. The mechanism has a paddle-wheel design with four blades, each with its own slit. The slits are identified by their projected widths on the Sun and are in the order 1\arcsec, 266\arcsec, 2\arcsec and 40\arcsec. The two wide slits are usually referred to as slots, and the present work focuses on the 40\arcsec\ slot, which we will often refer to as simply the slot.

EIS has two $2048\times 1024$ pixel CCDs, one for each of the two channels, with the long and short dimensions corresponding to wavelength and  the slit/slot length, respectively. The latter is aligned with the solar south-north direction. One pixel corresponds to 1.0\arcsec\ in the x and y dimensions and 22.3\,m\AA\ in the x direction (solar-x and wavelength dimensions are mixed for slot data).
The onboard software restricts downloaded detector images to 512~pixels in the y direction, so two exposures are required to obtain the complete extent of the EIS slits. Generally EIS observations are co-centered with the \emph{Solar Optical Telescope} on \emph{Hinode}, which results in a y-center located at about y-pixel 592 (80 pixels higher than the CCD center). Thus the y-pixel region most widely downloaded is 336 to 847. Y-pixels are indexed from 0 to 1023, although the onboard software prevents pixel 0 from being downloaded. The bottom pixel of the observed detector window is referred to as the y initial position (YIP).

The complete EIS wavelength bands are usually not downloaded in order to conserve telemetry. Instead \emph{wavelength windows} are specified, centered on strong or diagnostically-important emission lines. The onboard software restricts the window sizes to be multiples of eight pixels. Windows of 40 pixels are the minimum required for slot datasets, but 48 pixels are recommended to enable a background intensity level to be estimated.

\begin{figure}
    \centering
    \includegraphics[scale=0.65, angle=90]{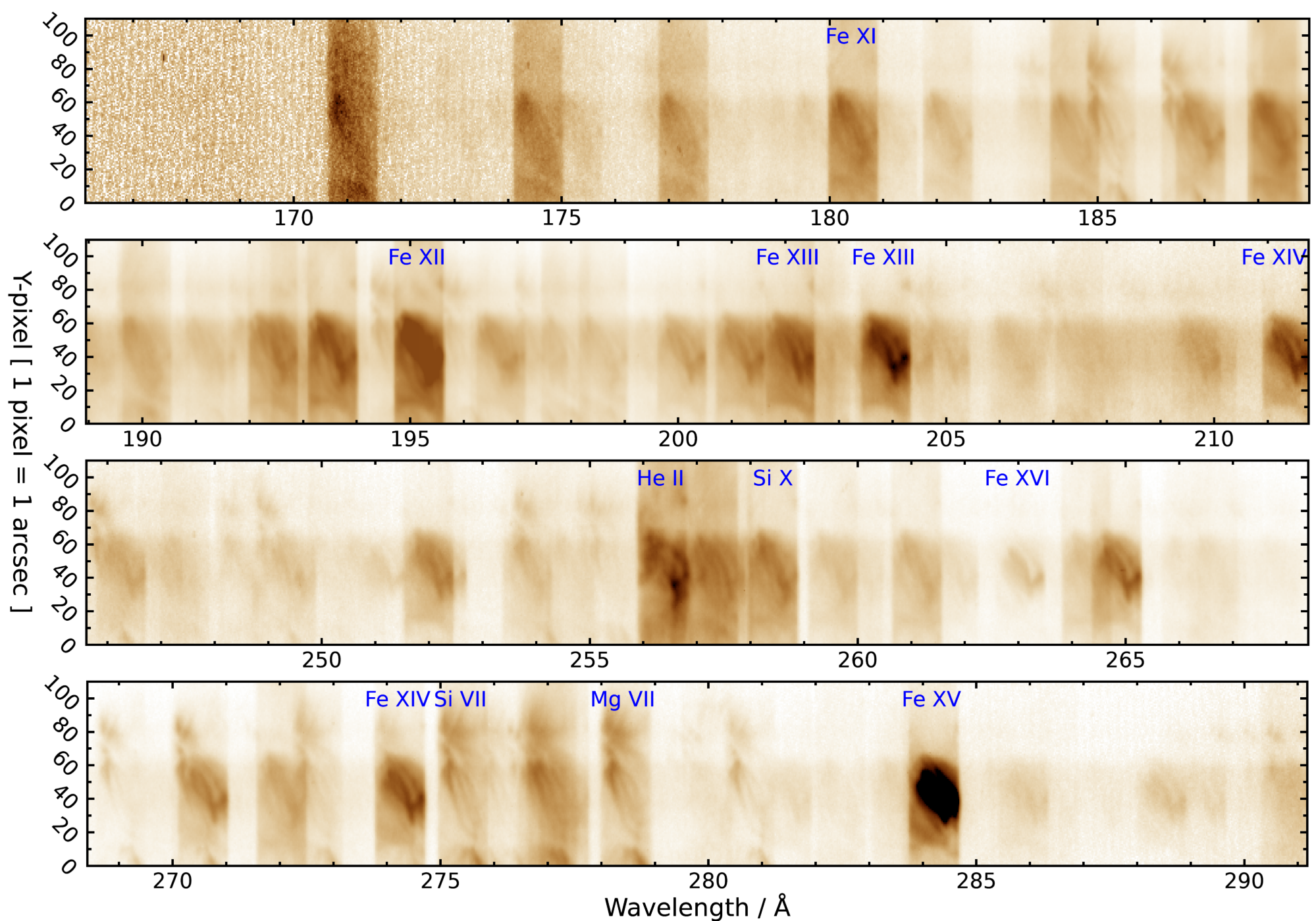}
    \caption{A complete EIS spectrum obtained with the 40\arcsec\ slot, derived from an observation on 15 May 2007 at 05:41\,UT. An inverse, square-root intensity scaling has been applied and an artificial saturation to bring out the weaker features.}
    \label{fig.all}
\end{figure}

Figure~\ref{fig.all} shows a slot exposure obtained on 15 May 2007 at 05:41\,UT with the EIS study SYNOP001. The complete EIS spectral range is shown over the four plots and 110 pixels along the y direction (corresponding to 110\,arcsec). A 30~s exposure was used. The radiometric calibration has been applied, which results in the \ion{Fe}{ix} \lam171.06 line appear quite bright even though the effective area at this location is very weak. Conversely, the strongest line in the EIS spectrum in terms of counts -- \ion{Fe}{xii} \lam195.12 -- appears less strong.
In the bottom half of the image there is a quiet Sun bright point that is most prominent in \ion{Fe}{xv} \lam284.16. The slot images as they appear on the detector are reversed in the X-direction compared to the true solar image.  Although the slot is identified as being 40\arcsec\ wide, it is actually closer to 41\arcsec\ \citep{2006ApOpt..45.8674K}. Due to the line spread function (LSF) of the instrument, the sharp edges of the slot are blurred in the images and the total extent of a slot image in the dispersion direction is about 46\arcsec.


The most prominent regular slot observation is the full-Sun scan performed every three weeks as part of \hinode\ Operation Plan (HOP) 130. The large-format study 480\as\ $\times$ 512\as\ \textsf{full\_sun\_slot\_scan\_2} (and its earlier version \textsf{full\_sun\_slot\_scan\_1}) is run over multiple spacecraft pointings in order to provide image mosaics  that cover the entire solar disk in a few hours. These data have been valuable for monitoring the EIS radiometric calibration \citep{2014ApJS..213...11W}. EIS \ion{Fe}{xii} \lam195.12 full-disk mosaics are compared with irradiance observations by the \textit{EUV Variability Experiment} \citep{2012SoPh..275..115W} to establish the EIS absolute calibration and the sensitivity decay over time.

All of the slot images are blended with images from nearby lines to some extent, but the lines listed in Table~\ref{tbl.lines} (and marked on Figure~\ref{fig.all}) give relatively clean images with a good signal. For each line in Table~\ref{tbl.lines} the wavelength from version~10 of CHIANTI \citep{2016JPhB...49g4009Y,2021ApJ...909...38D} is listed. The temperature of maximum emission [$T_\mathrm{mem}$] (computed with CHIANTI) is the temperature at which a line's contribution function peaks.

\begin{center}
\begin{table}[t]
\caption{Slot lines that are mostly unblended.}
\begin{tabular}{lccclcc}
\noalign{\hrule}
\noalign{\smallskip}
Ion & Wavelength & $\mathrm{Log}\,T_{\rm mem}$ & ~~ &
Ion & Wavelength & $\mathrm{Log}\,T_{\rm mem}$  \\
\noalign{\hrule}
\noalign{\smallskip}
\ion{He}{ii} & 256.32 & 4.95 & &\ion{Fe}{xiii} & 203.83 & 6.24\\
\ion{Mg}{vii} & 278.40 & 5.79 & &\ion{Fe}{xiii} & 202.04 & 6.25\\
\ion{Si}{vii} & 275.36 & 5.79 & &\ion{Fe}{xiv} & 211.32 & 6.29\\
\ion{Fe}{xi} & 180.40 & 6.12 & &\ion{Fe}{xiv} & 274.20 & 6.29\\
\ion{Si}{x} & 258.37 & 6.15 & &\ion{Fe}{xv} & 284.16 & 6.34\\
\ion{Fe}{xii} & 195.12 & 6.19 & &\ion{Fe}{xvi} & 262.98 & 6.43\\
\noalign{\hrule}
\end{tabular}
\label{tbl.lines}
\end{table}
\end{center}

\begin{figure}
    \centering
    \includegraphics[width=\textwidth]{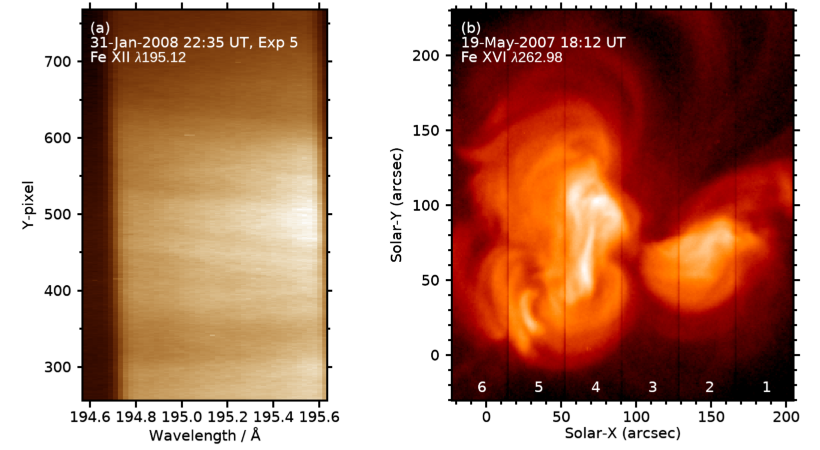}
    \caption{(a) a slot image of the Sun in \ion{Fe}{xii} \lam195.12 obtained in a quiet region above the west limb on 31 January 2008 at 22:35~UT. (b) a slot-raster image in the \ion{Fe}{xvi} \lam262.98 line, obtained from an observation on 19 May 2007 at 18:12\,UT. The numbers at the bottom indicate the order in which the raster is built from the six exposures. A logarithmic intensity scaling has been applied.}
    \label{fig.squash}
\end{figure}

Figure~\ref{fig.squash}(a) shows the \ion{Fe}{xii} \lam195.12 image from the fifth exposure of a raster observation obtained on 31 January 2008 at 22:35\,UT. It was obtained above the west limb of the Sun in quiet conditions and it is displayed in the detector frame, so east and west are reversed. The wavelength window was set to 48 pixels wide by 512 pixels high, and thus there is background emission on the left side, and a little on the right side. The aspect ratio has been modified  in order to better illustrate the tilt of the slot image, which can be seen to be about 2--3 pixels in the wavelength direction.

By adjusting the primary mirror position, wide-field images of the Sun can be obtained through raster scans and Figure~\ref{fig.squash}(b) shows an example from an active region observation on 19 May 2007 that used six raster positions. An exposure time of 30~s was used, and the step size between exposures was set to 36\arcsec. Thin vertical lines can be seen on the image, revealing the six individual exposures that comprise the total image. As will be shown later, the intensity falls off at the edges of the slot and so, even though the step size is smaller than the slot width, it is not small enough to avoid the intensity drop-offs. To obtain images with a completely smooth intensity distribution, a step size of 32~pixels is recommended, and this is implemented in HOP~130.

The EIS datasets used in this article are freely available from the \textit{Virtual Solar Observatory} (\urlurl{sdac.virtualsolar.org}), and the IDL software used for data analysis are available in the \textit{Solarsoft} (\urlurl{sohowww.nascom.nasa.gov/solarsoft}) library, except where otherwise noted.

\section{Slot Tilt, Width and Spatial Resolution}\label{sect.twsr}

In this section we address the first three questions identified in the Introduction by applying a fitting function to the intensity distribution across the dispersion direction of slot images. \ion{Fe}{xii} \lam195.12 is chosen for this analysis as it is normally the strongest line observed by EIS, and thus offers the best opportunity for constraining the slot parameters. The results enable us to determine the variation of slot width and spatial resolution along the full length of the slot, and to measure the tilt of the slot that is apparent in Figure~\ref{fig.squash}(a). The latter then allows us to compare the relative positions of the 1\as\ and 40\as\ slits on the detector.


\subsection{Fitting Method}\label{sect.fit}

\begin{figure}[t]
    \centering
    \includegraphics[width=\textwidth]{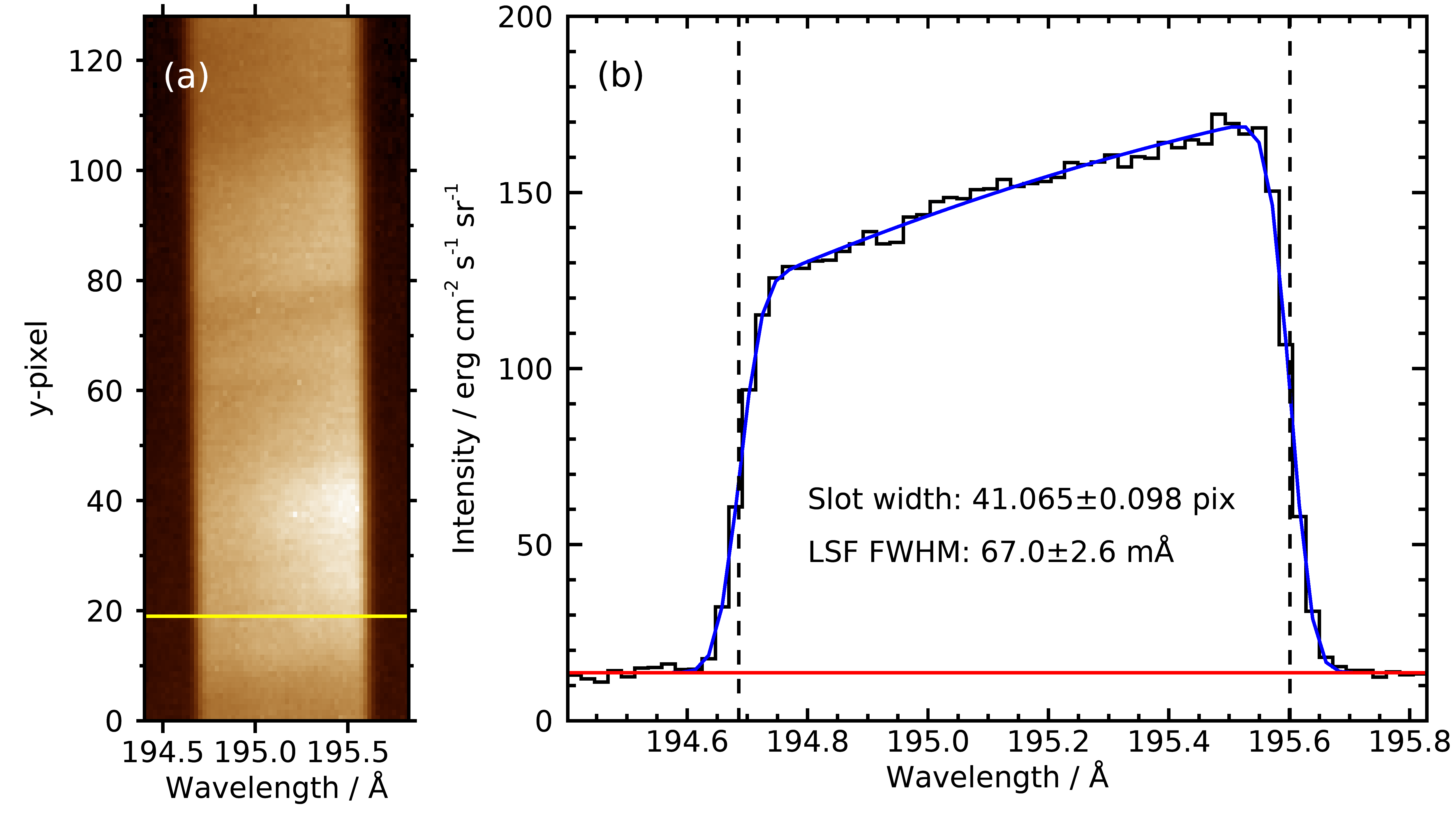}
    \caption{(a) the \ion{Fe}{xii} \lam195.12 slot image obtained at 13:12~UT on  20 September 2021. (b) the intensity cross-section through the slot image at y-pixel 19 (indicated by the \textit{yellow line} on Panel a). The \textit{blue line} shows the seven-parameter fit to the intensity, the \textit{horizontal red line} shows the fitted background intensity, and the \textit{vertical dashed lines} show the edges of the slot as derived from the fit.}
    \label{fig.fit}
\end{figure}

The method chosen for estimating the slot tilt and width was to perform a relatively simple parametric fit to the distribution of intensity in the dispersion direction (Figure~\ref{fig.fit}). \ion{Fe}{xii} \lam195.12 is ordinarily the strongest line observed by EIS and, because the corona emits strongly at the ion's formation temperature of 1.5\,MK, there is usually good signal over the full slot image frame. However, the emission is often highly inhomogeneous as illustrated in Figure~\ref{fig.all}, and thus not susceptible to a limited parameter fit. 

An exception is quiet Sun emission above the limb. Above heights of around 50\arcsec\ there is little contribution from bright points and the emission is relatively uniform with a decreasing intensity with radial distance. There is no option to roll the Hinode spacecraft so the EIS slot is always oriented north-south. The preferred observation for off-limb quiet Sun measurements is then for the slot to be centered near the Equator. 


Figure~\ref{fig.fit} shows an example from a dataset obtained on 20 September 2021 that is analyzed in the following section. The left panel shows the \ion{Fe}{xii} \lam195.12 slot image, and y-pixel 19 is indicated. (The data have been binned by 4 pixels in y to improve signal to noise; the original data have 512 y-pixels.) The intensity pattern across the slot is shown in the right panel, and a simple, almost linear decrease in intensity across the slot can be seen. 

The data for these off-limb datasets were fit as follows. The distribution of intensity shown in Figure~\ref{fig.fit}  was fit with a quadratic function, convolved with a boxcar function, and then convolved with a Gaussian. The boxcar is used to model the sharp edges of the slot, and the Gaussian simulates the line spread function of the instrument. The function is then added to a constant background. There are seven free parameters in this function: three parameters for the quadratic function, one for the left edge of the slot, one for the slot width, one for the width of the Gaussian, and one for the background. The IDL routine \textsf{eis\_fit\_slot\_exposure} was written to perform the fits and is available in a GitHub repository (\urlurl{github.com/pryoung/eis\_slot)}, where further details are given in a README file. The routine calls the MPFIT IDL procedures (\urlurl{purl.com/net/mpfit}) \citep{2009ASPC..411..251M} to obtain the best fit to the intensity profile through $\chi^2$ fitting. The fit for the intensity profile in Figure~\ref{fig.fit} is over-plotted as a \emph{blue line}, and the derived slot width and the full-width at half-maximum (FWHM) of the LSF are indicated. The LSF FWHM corresponds to 3.0\as, which agrees with the value of \citet{eis8} that was derived from images of small transition region brightenings.

\subsection{A Full-Slit Observation}\label{sect.sep20}


As discussed in Section~\ref{sect.overview}, emission from the slot extends over the full height of the detector but individual exposures are restricted to 512~pixels. It is thus necessary to schedule two observations for the bottom and top halves of the detector. The studies must also have a wavelength window of at least 48 pixels for \ion{Fe}{xii} \lam195.12 to enable  an accurate fit to the intensity profile. No such observations were found in the EIS archive and so a special observation was obtained on 20 September 2021.


Quiet Sun above the limb was observed and two pointings were performed for each raster to take the top and bottom halves of the detector. Multiple datasets were taken over an 80~min period beginning at 12:02~UT and we focus here on two rasters obtained with the \textsf{dei\_qs\_80\_slot40} study and beginning at 13:11:21\,UT and 13:12:28\,UT. The study performs a 2-step raster with a 40\arcsec\ step size and a 30\,s exposure time. Only the first steps of the two rasters are considered here, and the two exposures are centered at x=1080\as, about 120\as\ above the limb. The first exposure is at YIP=1 and the second at YIP=512.

\begin{figure}[t]
    \centering
    \includegraphics[width=\textwidth]{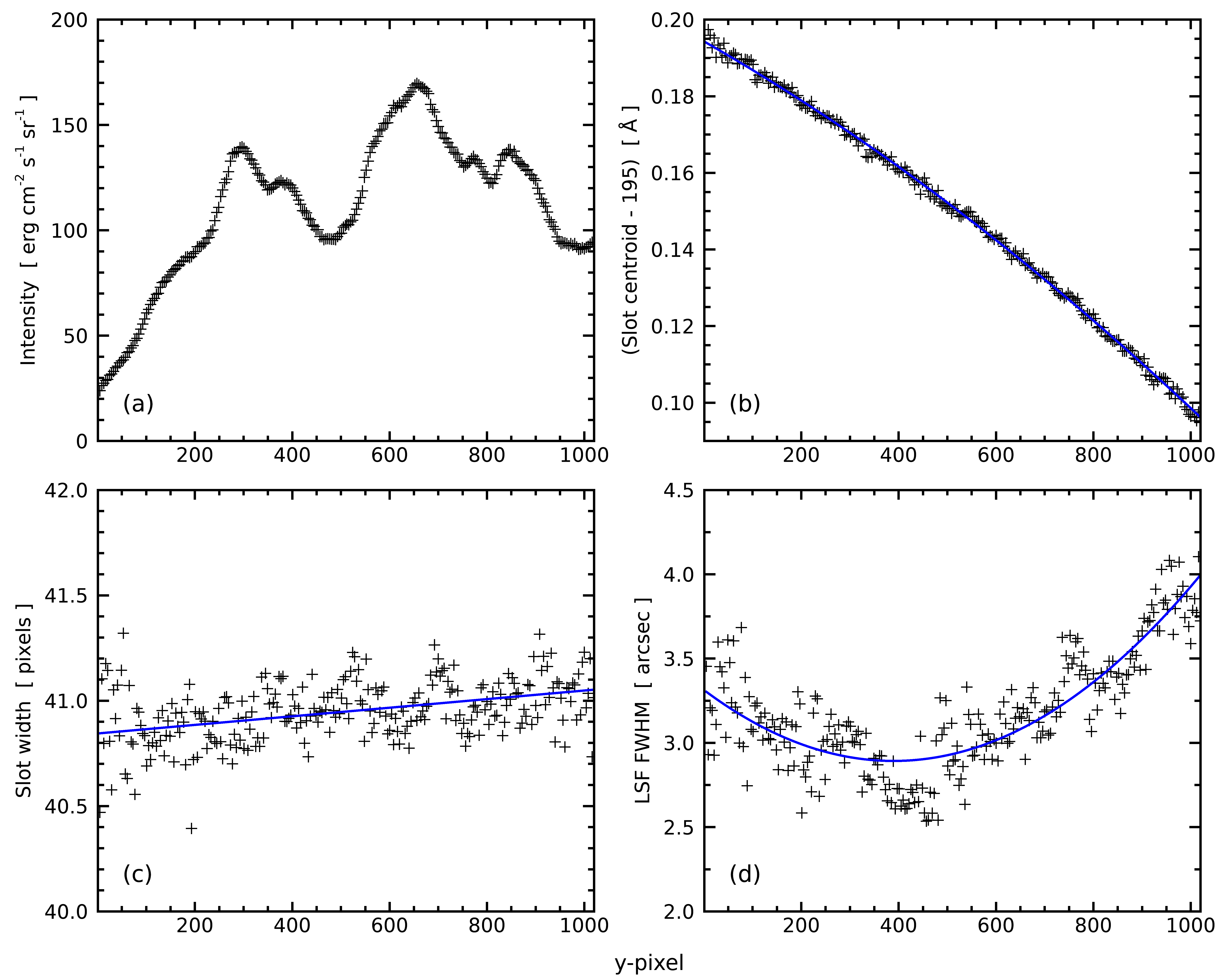}
    \caption{Slot parameters as a function of detector y-pixel for an observation obtained at x=1080\as\ on 20 September 2021. The parameters are: (a) average intensity across the slot, (b) slot centroid, (c) slot width, and (d) line spread function FWHM. The \textit{blue curves} on panels (b)--(d) show the polynomial fits to the data.}
    \label{fig.full}
\end{figure}

Using the fitting procedure described in the previous section, the \lam195.12 emission across the slot was fit with the seven-parameter function. Four-pixel binning was applied in the y-direction to reduce the impact of warm pixels. The bottom edge of the slot is apparent in the YIP=1 exposure, such that appreciable signal is only seen for detector y-pixel 7 and above. The top edge of the slot is not apparent in the YIP=512 exposure.

\begin{center}
\begin{table}[t]
\caption{Slot parameters derived from the 20 September 2021 observation.}
\begin{tabular}{cccc}
\hline
Quantity & $a_0$ & $a_1$ & $a_2$ \\
\hline
Centroid & -- & $-7.260(-5)$ & $-2.316(-8)$ \\
Width & 40.8445  & $2.038(-4)$ & --\\
LSF FWHM & $7.384(-2)$ & $-4.818(-5)$ & $6.186(-8)$ \\
\hline
\end{tabular}
\label{tbl.full-params}
\end{table}
\end{center}

Figure~\ref{fig.full} shows the results of the intensity profile fitting for the two datasets. The slot centroid is obtained from the fit parameters by adding half the slot width to the wavelength for the left edge of the slot. The average slot intensity shown in Panel (a) is computed by averaging the 35 pixels centered on the centroid pixel; the full slot width is not used as the intensity falls off near the edge of the slot (Figure~\ref{fig.fit}).  The slot centroid is shown in Panel (b), the slot width in Panel (c), and the FWHM of the Gaussian LSF is shown in Panel (d). The latter is plotted in arcsec units, and the conversion factor back to \AA\ is 0.02228. The three quantities were fit with polynomial functions: third-order functions for the centroid and LSF FWHM, and a second-order function for the slot width, which shows a much smaller variation. The fits are shown as \emph{blue curves} on Panels (b) to (d), and the polynomial coefficients are given in Table~\ref{tbl.full-params}. Thus the parameter value derived from the fits for a pixel $y$ is given by $\sum_i a_iy^i$. The $a_0$ value is not given for the slot centroid because the positions of lines on the detector shift during an orbit due to thermal effects \citep{2010SoPh..266..209K}, and so $a_0$ is dataset-specific.

The LSF FWHM is interpreted here as a measure of the spatial resolution of the slot. The best resolution of 2.9\as\ occurs at y-pixel 384, and the worst resolution of 4.0\as\ at the top of the detector (y-pixel 1023). Section~\ref{sect.overview} mentioned that EIS is usually centered on the SOT field-of-view, corresponding to y-pixel 592, where the resolution is 3.0\as.

\subsection{Time Variation}\label{sect.time}

The tilts of the narrow EIS slits were measured early in the mission and are obtained with the IDL routine \textsf{eis\_slit\_tilt}. Fourth-order polynomials were fit to the variation of line centroids over the full height of the detector, and the parameters are stored in the file \textsf{eis\_slit\_tilt.txt} within the EIS \textsf{Solarsoft} directory. An adjustment to the grating focus mechanism was performed on 24 August 2008 and so different parameters apply before and after this date. We thus expect a similar change for the slot parameters.

As slot data covering the full height of the detector, with a sufficiently wide spectral window, were not available prior to the 20 September 2021 observation described in the previous section, the best that can be done is to compare datasets that obtain the maximum 512 pixels allowed in one exposure.


\begin{center}
\begin{table}[t]
\caption{EIS datasets used to measure slot parameters.}
\begin{tabular}{ccccccccc}
\hline
Date & Time [UT] & Study & $N_\lambda$ & YIP \\
\hline
31-Jan-2008 & 22:37 & Slot\_raster &  48 & 256\\
5-Jun-2012  & 21:06 & SI\_Venus\_slot\_v1 & 48 & 319\\
9-May-2016  & 19:10 & SI\_Mercury\_slot\_v1 & 48 & 212 \\
13-May-2020 & 21:06 & Alignment\_modify\_v2 & 48 & 294 \\
\hline
\end{tabular}
\label{tbl.studies}
\end{table}
\end{center}

The EIS archive was searched for slot datasets with 512 pixels along the slot, at least 48 pixel wide wavelength windows, and a pointing above the limb in quiet Sun conditions. The spatial locations were checked using the \textsf{EIS\_Mapper} software (\urlurl{eismapper.pyoung.org}) to determine if the observations avoided active regions. Very few suitable datasets were found over the mission lifetime, but four were selected from different times during the mission and are listed in  Table~\ref{tbl.studies}. $N_\lambda$ is the number of wavelength pixels used for \lam195.12.

For each of the four datasets, one data file was chosen, and the best exposure within the raster was selected based on signal-to-noise and distance from the limb. Two-pixel binning was applied to each of the four datasets, and the intensity profiles were fit using the \textsf{eis\_fit\_slot\_exposure} routine. The slot centroids were derived from the fits, as described in Section~\ref{sect.fit}, and the variations of the centroids along the slit were fit with linear functions. The gradients are listed in the third column of Table~\ref{tbl.stud-results}. Linear functions were used instead of the quadratics used in Section~\ref{sect.sep20} because the datasets covered 512~pixels of the detector instead of 1024~pixels.

In order to compare with the full-slot data from Section~\ref{sect.sep20}, subsets of the full-slot centroid data were extracted to match the pixel ranges of each of the four datasets. Linear functions were then fit to these subsets to yield the tilt gradients given in column 4 of Table~\ref{tbl.stud-results}. The gradients vary because of the different YIP values of the four datasets (fifth column of Table~\ref{tbl.studies}).


As expected, there is a large difference between the 2008 dataset and the 2021 dataset, as the former occurred before the grating change on 24 August 2008. The tilt becomes smaller after this date, and the change is consistent with that found for the narrow slits. The 2012 and 2021 tilts are close, but the 2016 and 2020 datasets show larger tilts (by magnitude) than the 2021 dataset by 6\%. This could be due to thermal changes within the instrument.

The tilt is of value in the present work in relation to background subtraction for the slot intensities, and high accuracy is not required. Our recommendation is to use the Table~\ref{tbl.full-params} fit parameters for datasets after 24 August 2008, and the 31 January 2008 tilt value from Table~\ref{tbl.stud-results} before this date.

 
\begin{center}
\begin{table}[t]
\caption{Slot parameters derived from observations.}
\begin{tabular}{ccccc}
\hline
& & \multicolumn{2}{c}{Tilt [$\mu$\AA~pix$^{-1}$]} & \\
\cline{3-4}
Date & Exposure$^a$ & Measured & 2021 & Difference$^{b}$\\
\hline
31-Jan-2008 & 5/16 & $-113.0 \pm 0.8$ & $-93.7\pm 2.0$ & $-21.9$\% \\
5-Jun-2012 & 4/6 & $-99.7 \pm 0.6$  & $-98.9 \pm 2.0$ & $-0.8$\% \\
9-May-2016 & 3/6 & $-97.5 \pm 1.2$ & $-92.0 \pm 2.0$ & $-6.0$\% \\
13-May-2020 & 3/3 & $-102.2 \pm 0.5$ & $-96.3\pm 2.0$ & $-6.1$ \% \\
\hline
\multicolumn{5}{l}{$^a$ $m/n$ -- the $m$th exposure of $n$.}\\
\multicolumn{5}{l}{$^b$ percentage difference relative to the 2021 dataset.}\\
\end{tabular}
\label{tbl.stud-results}
\end{table}
\end{center}


\subsection{Slit and Slot Tilt Comparison}\label{sect.tilt-comp}

\begin{figure}[t]
    \centering
    \includegraphics[width=\textwidth]{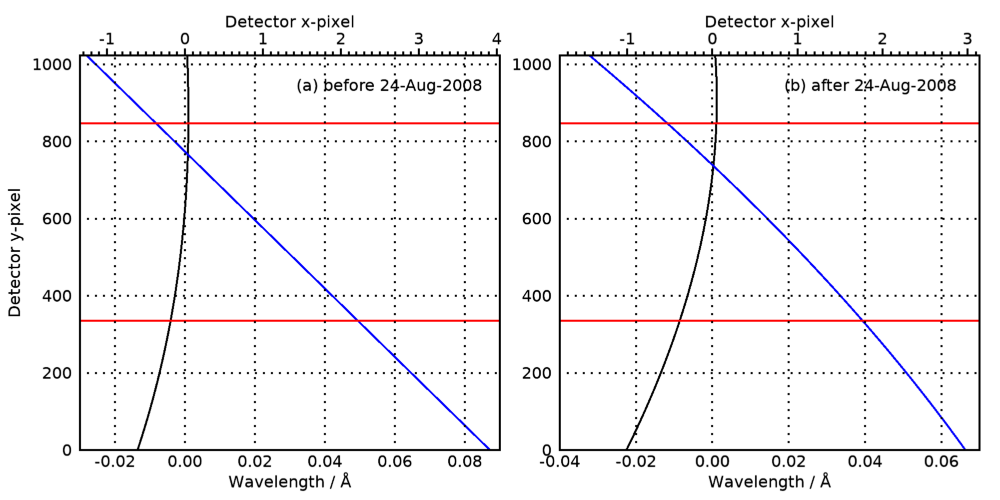}
    \caption{Comparisons of the 1\as\ slit (\textit{black}) and the centroid of the 40\as\ slit (\textit{blue}) as they appear on the EIS detectors. Panels (a) and (b) show the positions prior to and after the grating movement on 24 August 2008, respectively.
    The \textit{red horizontal lines} indicate the portion of detector that is most actively used.}
    \label{fig.slits}
\end{figure}

Figure~\ref{fig.slits} compares the position of the slot centroid on the EIS detector with that of the 1\arcsec\ slit for the periods before and after the grating change on 24 August 2008. The tilts of the slot are from the analysis of Sections~\ref{sect.sep20} and \ref{sect.time}. The slit tilts come from the IDL routine \textsf{eis\_slit\_tilt}. To obtain the relative positions of the slit and slot, it is necessary to use two exposures that are close in time, as the projected slit positions on the detector vary during the 98.5~min \hinode\ orbit. For the period before the grating movement, the SYNOP001 dataset beginning at 18:05~UT on 9 May 2007 was used. A patch of relatively uniform emission in the slot image was identified and the 7-parameter slot fit function was applied to yield the centroid. The same spatial region in the slit exposure was averaged and fit with a Gaussian to yield the slit centroid. The difference fixes the relative offset between the slot and slit. Further details on this process can be found in the code used to generate Figure~\ref{fig.slits}, available in the supplementary material at \urlurl{github.com/pryoung/papers/tree/main/2022\_eis\_slot}.

The 20 September 2021 dataset (Section~\ref{sect.sep20}) was used to derive the slot--slit offset for the period after the grating move. The \textsf{dei\_qs\_80\_slot40} exposure at 13:12:28~UT (used in Section~\ref{sect.sep20}) was followed by a slit exposure from the study \textsf{SK\_DEEP\_5x512\_SLIT1} at 13:14:50~UT. The latter study was run specifically to obtain an exposure at the same fine mirror position (see Section~\ref{sect.align}) as the preceding slot study and for the same detector y pixels. To match the processing of the slot data, the narrow slit data were binned in the y direction by four pixels, and \lam195.12 was fit with a Gaussian along the slit direction to yield the line centroids. The centroids were averaged over positions 48 to 52, and compared with the slot centroid at the same position to yield the relative offset of the slot to the slit.


Figure~\ref{fig.slits} shows that the  tilt is significantly greater for the slot, varying by more than four pixels over the detector height, compared to one pixel for the narrow slit. The curves intersect at y pixels 766 and 737 for the periods before and after the grating move, respectively. Thus a slot exposure is mostly offset to the right of a slit exposure on the detector. This is significant because the EIS planning software positions the slot wavelength windows assuming the centroid is the same as the 1\arcsec\ slit. This is why slot exposures such as the one shown in Figure~\ref{fig.squash}(a) are usually offset to the right of the wavelength window.






\section{Slit--Slot Intensity Comparison}\label{sect.int}


In this section we present the procedure for comparing \ion{Fe}{xii} \lam195.12 slot image intensities with intensities derived from the 1\as\ slit. Intensities are compared for a set of datasets obtained in May 2007, and a correction factor to be applied to slot data is suggested.


\subsection{Slit/slot Alignment}\label{sect.align}

EIS performs raster scans through fine-scale motions of the primary mirror. For a fixed mirror position, if the slit/slot mechanism is rotated from the 1\arcsec\ slit to the 40\arcsec\ slot, then the slit is approximately centered within the slot image. The user can check the fine mirror position [\textsf{FMIRR}] with the following IDL command
\begin{verbatim}
IDL> fmirr=*(d->getaux_data()).fmirr
\end{verbatim}
where \textsf{d} is an EIS data object.
For a raster observation there will be an \textsf{FMIRR} value for each step of the raster. 
A single step corresponds to about 0.25\arcsec\ but the EIS onboard software restricts mirror steps to 1\arcsec\ or 4 steps.
The EIS raster direction is always solar-west to solar-east, or right-to-left in displayed raster images. The raster direction corresponds to increasing \textsf{FMIRR} values.

\subsection{Intensity Calibration for Slot Data}

The routine \textsf{eis\_prep} converts the signal measured by the instrument to calibrated intensity units, and the steps involved are described in \citet{2022zndo...6339609Y}. One of the factors in this conversion is the effective area curve, which shows significant variation with wavelength (the curves can be generated with the IDL routine \textsf{eis\_eff\_area}). For example, the effective area takes values of 0.074, 0.30 and 0.033\,cm$^2$ at wavelengths 185, 195 and 205\,\AA, respectively.
Slot data differs from slit data in that the image in a strong line will dominate on the CCD over an extended wavelength region: 40 pixels in the case of the 40\arcsec\ slot and 266 pixels in the case of the 266\arcsec\ slot. Thus \ion{Fe}{xii} \lam195.12 extends over 194.7 to 195.6\,\AA\ in 40\arcsec\ slot data. In this case \textsf{eis\_prep} takes the effective area values at the edges of the wavelength window, averages them, and applies this value to the whole window. This is reasonable for small windows, such as those from the 40\arcsec\ slot, but not for the 266\arcsec\ windows. It is also not correct for datasets for which windows are the entire CCD. In the latter case, a single effective area value is applied to all wavelengths.
In these cases the solution is to leave the data in DN units by giving the \textsf{/noabs} keyword to \textsf{eis\_prep}. The window of interest (within the full CCD) should then be extracted and the calibration factor calculated for the wavelength that gives rise to the dominant emission. The routine \textsf{eis\_slot\_calib\_factor} is available in \textit{Solarsoft} for this purpose.


\subsection{The SYNOP001 Study}

In order to compare intensities obtained with the slot to those obtained with the slit, it is necessary to obtain exposures close in time to avoid any changes due to plasma evolution. For the present work, the EIS study SYNOP001 was used.

\textit{Hinode} is pointed to disk center every day in order for XRT to obtain full-disk synoptic images. Prior to February 2008 this was done approximately every six hours, but this was changed to every 12 hours following the failure of the X-band antenna and the corresponding reduction in telemetry. The XRT synoptic periods last 10~min and EIS typically performs short-duration studies. 

The study SYNOP001 was run during the XRT synoptic periods for the periods 2006 December through February 2008, and then replaced with SYNOP005, which ran until 2011 February. For the period February 2009 to October 2009  SYNOP005 and SYNOP001 were often alternated in the XRT synoptic slots.

SYNOP001 consists of two rasters. The first obtains two  exposures of 30~s and 90~s with the slot at the same spatial location, and the second obtains two exposures of 30~s and 90~s with the 1\arcsec\ slit. The default study setting uses the same FMIRR value for both rasters, which puts the 1\arcsec\  exposure approximately at the center of the slot image. The 90~s slot exposure often results in detector saturation for  \ion{Fe}{xii} \lam195.12, and is not used here. For the present work only the 30~s exposures are considered. The slot and slit 30~s exposures are separated in time by 2~min and 15~s.

SYNOP005 is a less data intensive study version of SYNOP001, with just single 90~s exposures for the slot and slit. Also 14 spectral windows were used rather than obtaining the full CCD. 

\subsection{Slit/slot intensity comparison from SYNOP001}\label{sect.slit-slot}

In this section we process the SYNOP001 datasets from May 2007 in order to compare the \ion{Fe}{xii} \lam195.12 slot and slit intensities for the same spatial locations on the Sun. As noted above, the slit exposure is obtained 2~min and 15~s after the slot exposure, which limits the effect of time evolution on the intensity comparison. 

There were 69 SYNOP001 slit-slot dataset pairs and, for each of these, a Gaussian was fit to the narrow slit \lam195.12 profile at each of the 256 pixels along the slit using the IDL routine \textsf{eis\_auto\_fit} \citep{2022zndo...6339584Y}. The average centroid of the fits was then matched to the nearest wavelength pixel on the CCD. The same pixel in the slot dataset was identified, and the intensity was averaged for this pixel and its neighbors, to allow for the small tilt of the narrow slit. 

The Gaussian fit to the slit data automatically accounts for the background intensity level in the narrow slit spectrum. For the slot, the background  intensity was computed by first obtaining the centroid pixel of the slot by adjusting the slit centroid for the slot--slit offset (Figure~\ref{fig.slits}). The two pixels $\pm 28$~pixels from the slot centroid were identified and, for each, the five pixels centered on these pixels were averaged, yielding background estimates on the left and right sides of the \lam195.12 slot image. These two values were then averaged to give the final background intensity.

\begin{figure}[t]
    \centering
    \includegraphics[width=\textwidth]{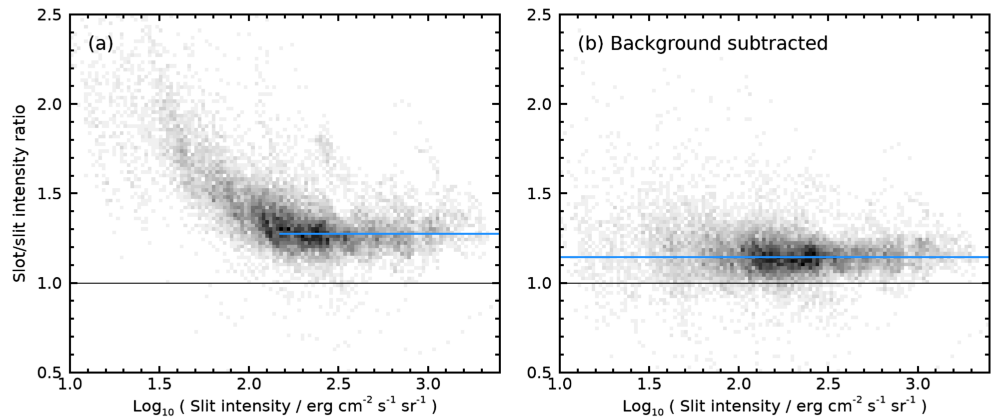}
    \caption{Heatmaps showing ratios of slot to slit intensities for the May 2007 SYNOP001 datasets, plotted as a function of the slit intensity. The bin sizes are 0.02 in both dimensions, and the maximum values are 53 (a) and 56 (b). For display purposes the maps are shown scaled to the power 0.7. Panel (b) shows the effect of background subtraction for the slot intensities. The \textit{blue horizontal line} on Panel (a) shows the mean ratio value for slit intensities $\ge$~150~\ecss, and that on Panel (b) shows the mean ratio value for all slit intensities.}
    \label{fig.slit-slot}
\end{figure}

Two of the 69 datasets were rejected due to anomalies with the slot image, and four further datasets because either the YIP or FMIRR values did not match between the slit-slot pairs. In addition, all slot intensities larger than 2740~\ecss\ were rejected as the detector saturates at this level. For all remaining pixels, Figure~\ref{fig.slit-slot}(a) shows the ratios of the slot to slit intensities, while Figure~\ref{fig.slit-slot}(b) shows the ratios after subtracting the slot background. This demonstrates clearly that background subtraction is necessary for slit intensities $\le$~150~\ecss. For intensities larger than this, the ratios without background subtraction are approximately constant and we find an average ratio of  1.27 with a standard deviation of 0.13. For the entire set of background-subtracted data the average ratio is 1.14 with a standard deviation of 0.14, and for intensities $\ge$~150~\ecss\ the average ratio and standard deviation are 1.13 and 0.10. For reference we note that 
\citet{2009ApJ...705.1522B} gave an average intensity for \lam195.12 in quiet Sun of 135~\ecss. 

We do not have an explanation for why the slot intensity is systematically higher than the slit intensity across the SYNOP001 datasets. We do not believe that it is due to blending not accounted for by the background subtraction. For example, we know that about half of the image in the weak \ion{Fe}{viii} 194.66~\AA\ line overlaps with the \lam195.12 image. The \ion{Fe}{viii} image will be inhomogeneous, and thus the five-pixel strip used to define the background on this side of the \lam195.12 image and which runs through the \lam194.66 image will not necessarily represent that part of the \lam194.66 image that overlaps with \lam195.12. If the background strip is weaker than the overlap region then the derived \lam195.12 intensity will end up brighter due to the unaccounted for \ion{Fe}{viii} component. But, equally, if the background strip is brighter than the overlap region, then the \lam195.12 will be lower than expected. Thus, on average, we would expect inhomogeneities in the blending line images to cancel out rather than lead to an over-estimate of the \lam195.12 intensity.


Without an explanation for the enhanced slot intensities, our suggestion is that \ion{Fe}{xii} \lam195.12 slot intensities should be divided by the  \textit{empirical correction factor} of 1.14 in order to reproduce the 1\as\ slit intensities.

The fact that the non-subtracted intensity ratio becomes constant for higher intensities implies that the background level in the slot spectrum scales with the \ion{Fe}{xii} intensity. This is despite the fact that the emission lines contributing to the background are not due to \ion{Fe}{xii}. Our explanation is that, although lines formed at different temperatures vary in different amounts according to solar conditions (coronal hole, quiet Sun, active region), to a first approximation they scale together. Consider the intensity measurements from Table~2 of  \citet{2008ApJS..176..511B}, in particular the ``QS", ``AR2" and ``Limb" values for the nearby lines of \ion{Fe}{viii} \lam194.66, \ion{Fe}{xii} \lam195.12 and \ion{Fe}{x} \lam195.40. Normalizing the \lam195.12 intensity to 100 in each of the three regions, the intensities for \lam194.66 are 10.8, 11.0 and 6.2 for QS, AR2 and Limb. The intensities for \lam195.40 on the same scale are 6.4, 6.9 and 4.0. That is, to within a factor two, the \ion{Fe}{viii} and \ion{Fe}{x} intensities scale with \ion{Fe}{xii} for these three regions, even though \ion{Fe}{xii} varies by more than a factor 10 across the three regions.

This scaling of the background with \ion{Fe}{xii} intensity will prove valuable later in providing a prescription for deriving intensities from slot data (Section~\ref{sect.40pix}).



\section{A Prescription for Measuring Intensities from Slot Data}\label{sect.pres}

The results from the present work suggest the following prescription for measuring accurate \ion{Fe}{xii} \lam195.12 line intensities from EIS slot data when the wavelength windows are at least 48 pixels wide:
\begin{enumerate}
\item Calibrate the data to \ecss\ units.
    \item Estimate the background level along the slot length (Section~\ref{sect.bg}). 
    \item Subtract the y-dependent background intensity from the slot image intensity.
    \item Divide the subtracted slot image intensity by 1.14.
    \item Add an uncertainty of 10\%\  in quadrature to the uncertainty provided by \textsf{eis\_prep}, to reflect the uncertainty in the derived slot-to-slit intensity ratios (Sect.~\ref{sect.slit-slot}).
\end{enumerate}
The resulting slot intensities should then be comparable with the slit intensities.

Section~\ref{sect.40pix} discusses the procedure to be applied for datasets that have 40-pixel windows for \lam195.12.

\subsection{Background Subtraction}\label{sect.bg}

Section~\ref{sect.slit-slot} provided a method for calculating the background emission using 5-pixel wide regions located 28 pixels either side of the slot centroid. This was possible because the SYNOP001 study returns the entire EIS wavelength range. Of the 88 slot rasters in the EIS database, 41 use a 40-pixel window for \lam195.12, 27 use a 48-pixel window, and the remainder use a window that is 56-pixels or larger. For the latter, the windows are wide enough to give access to background either side of the slot image. 

As noted earlier, slot images extend over 46 pixels on the detector, and so there is no background solution for  40-pixel windows (see the following section). 
For 48-pixel window data, however, it is possible to take advantage of the fact that the slot image is offset to the right-side of the wavelength window as highlighted in Section~\ref{sect.tilt-comp} and seen visually in Figure~\ref{fig.squash}(a). We can thus use the leftmost column of pixels in the slot image to define the background. 



\begin{figure}[t]
    \centering
    \includegraphics[width=\textwidth]{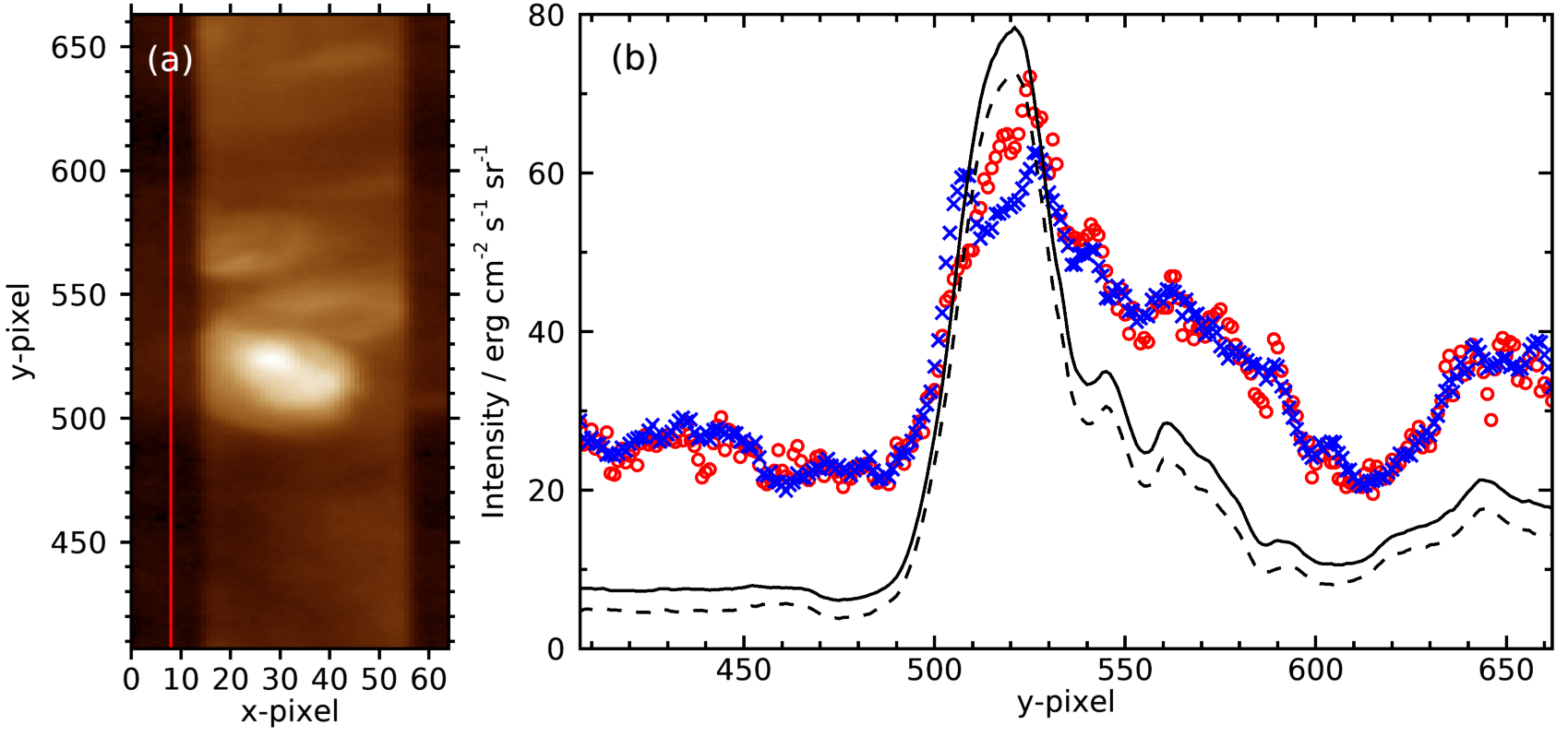}
    \caption{Panel (a) shows the \ion{Fe}{xii} \lam195.12 slot image from the 16 May 2007 10:46\,UT dataset, with a square-root intensity scaling. The \emph{vertical line} shows the location of the pixel that would correspond to the leftmost pixel of a 48-pixel wide window centered on \ion{Fe}{xii} \lam195.12. The intensity as a function of detector y-pixel position for this pixel is shown as \emph{red circles} on Panel (b). The \emph{blue crosses} show the average intensity for an alternative background prescription (see main text). The \emph{black solid line} shows the \ion{Fe}{xii} \lam195.12 intensity averaged over the 41 pixel width of the slot and divided by a factor 10 for display purposes. The \emph{black dashed line} shows the \lam195.12 intensity variation after background subtraction, also divided by 10.}
    \label{fig.bg}
\end{figure}

To test this method we selected the SYNOP001 dataset beginning at  10:46\,UT on 16 May, which shows a mixture of dark and bright features in the image (Figure~\ref{fig.bg}a). The intensity average over the 41-pixel width of the slit as a function of y-pixel is shown in Figure~\ref{fig.bg}(b) as a \textit{solid black line}. The \textit{blue crosses} on this panel show the background level derived using the method described in Section~\ref{sect.slit-slot}. The \textit{red vertical line} on Figure~\ref{fig.bg}(a) shows the location of the leftmost pixel if a 48-pixel window had been used for the observation. The intensity along this line is shown as \textit{red circles} on Panel (b). Excellent agreement is found with the \textit{blue crosses}, and the standard deviation of the differences between the two background methods is 7.6\,\%. This demonstrates that the 48-pixel window background subtraction method  gives good results.

For this particular dataset the slot image is clearly offset to the right of the wavelength window, and so the method of choosing the leftmost column for the background is successful. Figure~\ref{fig.slits} shows that the slot image tilts over to the left side of the wavelength window for y-pixels 800 and higher. In addition, there is a 1.5 pixel shift of the slot image during an orbit due to thermal effects that can lead to the slot image being more towards the left of the wavelength window depending when the exposure is taken. Therefore our general recommendation for estimating the background from 48-pixel window data is to take the leftmost and rightmost columns of the wavelength window and set the background for a specific row on the detector to be the minimum of these two values. 




In summary, for \lam195.12 windows of 56 pixels or more it is recommended that the background is calculated by averaging regions either side of the slot image. These regions can either be one or more data columns, or one can select regions that are equidistant from the slot centroid, as done in Section~\ref{sect.slit-slot}. For 48-pixel windows, the background can be obtained by taking the minimum value of the left and rightmost columns, as described above. A background estimate is not possible for 40-pixel windows.

\subsection{Datasets With 40-pixel Windows}\label{sect.40pix}

The previous section noted that 41 of the 88 EIS studies that use the slot use 40-pixel wavelength windows for observing \ion{Fe}{xii} \lam195.12. Our prescription for computing intensities from the slot data will therefore not work on these datasets since it is not possible to accurately estimate the background intensity level.
In Section~\ref{sect.slit-slot} it was noted that the ratio of the slot intensity without background subtraction to the slit intensity becomes constant above slit intensities of 150~\ecss. This suggests the following prescription for 40-pixel data:
\begin{enumerate}
    \item Calibrate the data to \ecss\ units.
    \item Divide the slot intensity by a factor 1.27.
    \item Disregard those pixels with intensities less than 150~\ecss. 
    \item Add an uncertainty of 10\%\  in quadrature to the uncertainty provided by \textsf{eis\_prep}, to reflect the uncertainty in the derived slot-to-slit intensity ratios (Sect.~\ref{sect.slit-slot}).
\end{enumerate}
In particular this should yield reliable \lam195.12 intensities for active regions.

\subsection{Extension to Other Emission Lines}

Our prescription for correcting the slot intensities applies only to \ion{Fe}{xii} \lam195.12, and here we discuss whether the method can be extended to other emission lines. 
It should work for any line that dominates over its immediate neighbors such that the background can be estimated, either on both sides of the slot image, or on the left side of the image. Referring to Figure~\ref{fig.all}, such lines include \ion{Fe}{xi} \lam180.40, \ion{Fe}{xiii} \lam202.04, \ion{Fe}{xiii} \lam203.82, \ion{Si}{x} \lam258.37, \ion{Fe}{xvi} \lam262.99, \ion{Fe}{xiv} \lam274.20, \ion{Si}{vii} \lam275.36, \ion{Mg}{viii} \lam278.40 and \ion{Fe}{xv} \lam284.16. For each of these it  will be necessary to repeat the analysis of Figure~\ref{fig.slit-slot} to check if the empirical correction factor is the same and also to determine the cutoff intensity in the case that 40~pixel windows are to be analyzed. As the SYNOP001 study returns the complete EIS wavelength range, then the May 2007 datasets are suitable for each of these additional lines.

\section{Summary}\label{sect.summary}

The Introduction to this article asked four questions in relation EIS 40\as\ slot data, and the answers are listed here.
\begin{enumerate}
\item \textit{What is the measured width of the slot?} The  width of the slot on the detector shows a small, linear increase from the bottom of the CCD to the top (Figure~\ref{fig.full}c) with a mean value of 40.949\as.
\item \textit{What is the tilt of the slot on the detector?} The position of the slot image on the detector varies quadratically with detector-y pixel (Figure~\ref{fig.full}b) with parameters given in Table~\ref{tbl.full-params}. The image shifts towards shorter wavelengths with increasing y-position, with a total shift of more than 4 pixels from the bottom of the CCD to the top. It was not possible to fully describe the slot tilt for data obtained prior to the EIS grating adjustment of 24 August 2008 and so only a linear fit to the tilt was performed and the gradient is given in Table~\ref{tbl.stud-results} (31 January 2008 dataset).
\item \textit{What is the spatial resolution of the slot?} The spatial resolution is measured from the fall-off of intensity at the edge of the slot image, and it varies quadratically with y-position (Figure~\ref{fig.full}d), with parameters given in Table~\ref{tbl.full-params}. The best resolution is 2.9\as.
\item \textit{Are the intensities measured with the slot compatible with the narrow slits?} Intensities measured with the slot are around 14\%\ higher than those measured with the 1\as\ slit, and it is recommended to reduce them by this amount. The background intensity in slot data is significant for intensities typical of quiet Sun or coronal hole regions and should  be subtracted. A prescription for deriving intensities from the slot data is given in Section~\ref{sect.pres}. 
\end{enumerate}


The results presented here should prove valuable to other researchers looking to derive quantitative results from the EIS slot data and exploit the large archive of observations obtained since 2006.

\begin{ack}
\hinode\ is a Japanese mission developed
and launched by ISAS/JAXA, with NAOJ as domestic
partner and NASA and STFC (UK) as international partners.
It is operated by these agencies in co-operation with ESA and
NSC (Norway). We thank Enrico Landi for planning the observations described in Section~\ref{sect.sep20}.
\end{ack}

\begin{fundinginformation}
P.R.~Young acknowledges support from  the  GSFC Internal Scientist Funding Model competitive work package program and the Heliophysics Guest Investigator program. I.~Ugarte Urra was funded by NASA under a contract to the U.S.\
Naval Research Laboratory.  
\end{fundinginformation}

\begin{dataavailability}
The level-0 EIS data files used for this work are publicly available from the \textit{Virtual Solar Observatory} (\urlurl{virtualsolar.org}) and the \textit{Hinode Science Data Center Europe} (\urlurl{sdc.uio.no}). Data derived as part of the analysis for the article are available from the \textsf{GitHub} repository \urlurl{github.com/pryoung/papers/tree/main/2022\_eis\_slot}.
\end{dataavailability}

\begin{ethics}
\begin{conflict}
The authors declare that they have no conflicts of interest.
\end{conflict}
\end{ethics}

\bibliographystyle{spr-mp-sola}
\bibliography{ms}{}

%

%

\end{article} 
\end{document}